\documentclass{aa}
\include{epsf}

\begin{document}

\thesaurus{6(02.03.3; 02.13.2; 06.01.2; 06.13.1; 06.18.2, 08.13.1)}

\title{On the transport of magnetic fields by solar-like stratified convection}

\author{S.B.F. Dorch\inst{1}
\and {\AA}. Nordlund\inst{2} }

\institute{The Royal Swedish Academy of Sciences, Stockholm Observatory,
           SE-13336 Saltsj\"{o}baden, Sweden (dorch@astro.su.se)
\and The Niels Bohr Institute for Astronomy, Physics and Geophysics,
     Juliane Maries Vej 30, DK-2100 Copenhagen {\O}, Denmark (aake@astro.ku.dk)}

\offprints{S.B.F. Dorch}

\date{Received date, accepted date}

\authorrunning{Dorch \& Nordlund}
\maketitle

\begin{abstract}
The interaction of magnetic fields and stratified convection was
studied in the context of the solar and late type stellar dynamos by
using numerical 3D MHD simulations. The topology
of stratified asymmetric and over-turning convection enables a pumping
mechanism that may render the magnetic flux storage problem obsolete.
The inclusion of open boundary conditions
leads to a considerable flux
loss unless the open boundary is placed close to the physical boundary.
Simulations including solar-like latitudinal shear indicates that a toroidal
field of several tens of kilo-Gauss may be held down by the pumping
mechanism.
\keywords{Sun: MHD -- magnetic fields -- stratified convection -- solar dynamo:
Stars -- magnetic fields}
\end{abstract}

\section{Introduction}

Magnetic fields play an important r\^{o}le for the formation of the
spectra of active late type G, K and M dwarf stars (e.g.
Schrijver \& Rutten \cite{Schrijver+Rutten87}, Rutten et al.\
\cite{Rutten+ea89} and Johns-Krull \& Valenti
\cite{Johns-Krull+Valenti96}). The most studied star in that respect
is the Sun.
It is generally assumed that solar active regions are manifestations
of a strong
toroidal flux system, that is generated and anchored deep below the surface
of the Sun, possibly in the undershoot layer below the convection zone proper.
Toroidal magnetic strands ascend through the convection
zone because of buoyancy and the on the average
super-adiabatic stratification.
The much weaker poloidal field is assumed to be
generated in the convection zone from this toroidal field
by a cyclonic effect.

In the mean field dynamo context, the mechanism that takes care of communicating
the poloidal field back to the region where the toroidal field is generated,
is assumed to be a diffusive coupling of the regions where the $\omega$-effect
and the cyclonic $\alpha$-effect operate (e.g.\ Parker \cite{Parker1993}).
Choudhuri \& Dikpati (\cite{Choudhuri+ea95})
have shown that meridional circulation may also couple the
two regions, if the time scale of the circulation is shorter than the
diffusive time scale, and if the circulation is such that the flow is
equator-ward at the bottom of the convection zone.
The modeling by diffusive coupling of the regions may be considered to be
somewhat unsatisfactory
because it relies on a rather ad hoc approach and
the meridional circulation approach is equally
unsafe since it is not well observed --- at best the
amplitudes and directions of the flow are indicated
(van Ballegooijen \cite{Ballegooijen1998}).

The question of how the poloidal field may return to the region where
the generation of the toroidal field
supposedly takes place is related to the
well-known ``buoyancy dilemma'' and the generally assumed solution to
this problem: Any magnetic field in the convection zone will escape
because of its buoyancy but a magnetic field may be stored in the stably
stratified region below the convection zone;
i.e., in the undershoot layer.
In that scenario the convection zone
is considered to be a passive one-dimensional medium while the magnetic
fields are treated more or less as
solid objects that move in it. It is conceivable, however, that
some kind of balance may occur between the drag of descending plasma
and the buoyancy of the magnetic field embedded in the plasma.

Along this line of thinking
Drobyshevski \& Yuferev (\cite{Drobyshevski+ea74})
proposed that a downward ``topological pumping''
of the magnetic field could be occurring, because of the asymmetric nature
of the topology of 3D convective flows, i.e.\ that they consist of
networks of descending material embedding regions of ascending material.
They investigated the kinematic case at low magnetic Reynolds number by
assuming an incompressible simple geometrical velocity flow pattern.
Criticism by Parker (\cite{Parker1975a}) made Drobyshevski et al.
(\cite{Drobyshevski+ea80})
redo the experiments with a more suitable upper boundary condition.
Arter et al. (\cite{Arter+ea82}), Arter (\cite{Arter1983}) and
Galloway \& Proctor (\cite{Galloway+ea83})
extended the
work to higher magnetic Reynolds numbers and several different
compressible flows.
They found that the magnetic energy did indeed increase at the bottom of
the domain.

During the last decade several groups have performed more detailed
magneto-convective numerical simulations (e.g.\
Hurlburt \& Toomre \cite{Hurlburt+ea88};
Brandenburg et al. \cite{Brandenburg+ea90};
Jennings et al. \cite{Jennings+ea92};
Nordlund et al. \cite{Nordlund+92};
Hurlburt et al. \cite{Hurlburt+ea94};
Nordlund et al. \cite{Nordlund+94a};
Brandenburg et al. \cite{Brandenburg+96jfm}).
Moreover, numerical simulations of stratified convection
have shown that trace particles initially placed in a horizontal layer
of a highly stratified model on the average are transported
downwards, as a result of the asymmetric topology of stratified convection
(Stein \& Nordlund \cite{Stein+Nordlund89b}).
In the context of magnetic fields such a tendency for downwards transport
of magnetic fields has also been seen
in convective dynamo simulations (Nordlund et al. \cite{Nordlund+92};
Brandenburg et al. \cite{Brandenburg+96jfm}).
Most recently
Tobias et al. (\cite{Tobias+ea98})
have investigated this effect using their standard
method for studying stellar convection and Mcleod (\cite{Mcleod98})
presented speculations along the same line of thought
that shall be followed here.

\section{Model}

The objective of the numerical experiments presented here is to study the
interaction of magnetic fields and solar-like stratified over-turning
convection and differential rotation.
The model of the Sun is a ``local box'' model of a convectively
unstable layer (henceforth referred to as the ``convection zone'')
sandwiched between two stable layers.
In order to circumvent problems associated with the very disparate
thermal and dynamical time scales, the model has a much
higher luminosity than the Sun, and all variables are
scaled accordingly.
To compare with solar values, the results must be re-scaled as follows.
With a flux scale of
f$_{\rm scl} = 3\times 10^5$ times the solar flux, the velocity scale factor
becomes u$_{\rm scl} = {\rm f}_{\rm scl}^{1/3} \sim 67$, the temperature
fluctuation
scale factor t$_{\rm scl} = {{\rm u}_{\rm scl}}^2 \sim 4.5\times 10^3$,
and the magnetic
field strength scale factor b$_{\rm scl} =$ u$_{\rm scl}$.
For convenience, the variables in the model are given in units of
time $u_{\rm time} = 10^3$ s, length $u_{\rm length} = 1$ Mm, and
density $u_{\rho}=1$ gcm$^{-3}$.
A Kramers' opacity scaled inversely with the scale factor
for the total luminosity is adopted.
This ensures that the boundary between the
stable layer and the convection zone is at about the same depth
in the model as it is in the Sun.

\subsection{Equations}

The full resistive and compressible MHD-equations are solved
using the staggered mesh
method by Galsgaard and others (e.g.\
Galsgaard \& Nordlund \cite{Galsgaard+ea97},
Nordlund \& Stein \cite{Stein+Nordlund89b},
\cite{Nordlund+92} and \cite{Nordlund+94a}):
\begin{eqnarray}
 \frac{\partial \rho}{\partial t} &
             = & - \nabla \cdot \rho {\bf u},\\ \label{mass}
 \frac{\partial \rho {\bf u}}{\partial t} &
             = & - \nabla \cdot ( \rho {\bf u}{\bf u} - {\bf \tau})
                 - \nabla P + {\bf F}_g + {\bf F}_{\rm Lorentz}
                 + {\bf F}_{\rm ext}, \label{EoM}\\
 \frac{\partial {\bf B}}{\partial t} &
             = & - \nabla\times {\bf E},\\ \label{field}
 \mu_0 {\bf j} &
             = & \nabla \times {\bf B},\\
 {\bf E} &
             = & \eta {\bf j} - {\bf u} \times {\bf B},\\
 \frac{\partial e}{\partial t} &
             = & - \nabla \cdot (e {\bf u}) + P(\nabla \cdot {\bf u})
                 + Q_{\rm rad} + Q_{\rm visc} + Q_{\rm Joule}, \label{energy}
\end{eqnarray}
where $\rho$ is the mass density, ${\bf u}$ the velocity, $\tau$ the viscous
stress tensor, $P$ the gas pressure, ${\bf B}$ the magnetic field density,
${\bf j}$ the current density, ${\bf E}$ the electric field,  $\eta$ the
magnetic diffusivity, and $\mu_0$ is the magnetic vacuum permeability.
In the momentum equation (\ref{EoM})
${\bf F}_g = \rho {\bf g}$ is the gravitational
force, ${\bf F}_{\rm Lorentz} = {\bf j} \times {\bf B}$ is the Lorentz force
and ${\bf F}_{\rm ext}$ is the sum of other forces
associated with the rotation to be discussed below (Sect.\ \ref{lat.shear}).
The gravitational acceleration ${\bf g}$ is along the x-direction
(equivalent to the radial direction).
Furthermore $e$ is the internal thermal energy,
$Q_{\rm visc}$ is the viscous heating, $Q_{\rm Joule}$ is the
Joule heating,
and $Q_{\rm rad}$ is the radiative heating (cooling).

The code uses a finite difference staggered mesh with 6th order derivative
operators, 5th order centering operators and a 3rd order time-stepping
routine (Hyman \cite{Hyman1979}).

\subsection{Boundary conditions}

In most stars envelope convection is essentially driven
by surface cooling. The entropy contrast at the surface is far
larger than that at the bottom of the convection zone, if the convection
zone extends over several or many pressure scale heights.  To model
this situation, without having to actually include all layers up to the
solar surface, a simple
expression for an isothermal cooling layer at the upper boundary of the model
was used:
\begin{equation}
Q_{\rm rad} = - \frac{( T - T_{\rm top} )}{\tau_{\rm cool}} f(x),
\end{equation}
where $\tau_{\rm cool}$ is the characteristic cooling time,
$T_{\rm top}$ is the temperature of the cooling layer,
and $f(x)$ is a profile function that restricts the effect to a thin
surface layer.

Both experiments with closed and open upper boundaries were performed.
In order to implement a stable open upper boundary
a buffer zone was allocated, where a fiducial electric field is
gradually turned on.
The sense of the electric field is such that it
drags the magnetic field out of the buffer zone, and the magnitude is
increased from zero to of the order of $u_{\rm max} B$, where $u_{\rm max}$
is the maximum velocity in the buffer zone.
A layer that is about 10 grid zones from the numerical upper
boundary may thus be considered as the physical open upper boundary.
This layer is far below the real boundary of the solar convection zone
(even the numerical upper boundary of the model is far below the
photosphere).

\begin{figure*}[!htb]
\centering
\makebox[14cm]{
\vspace{0cm}
\hbox{\hspace{0cm}\epsfxsize=7.0cm \epsfbox{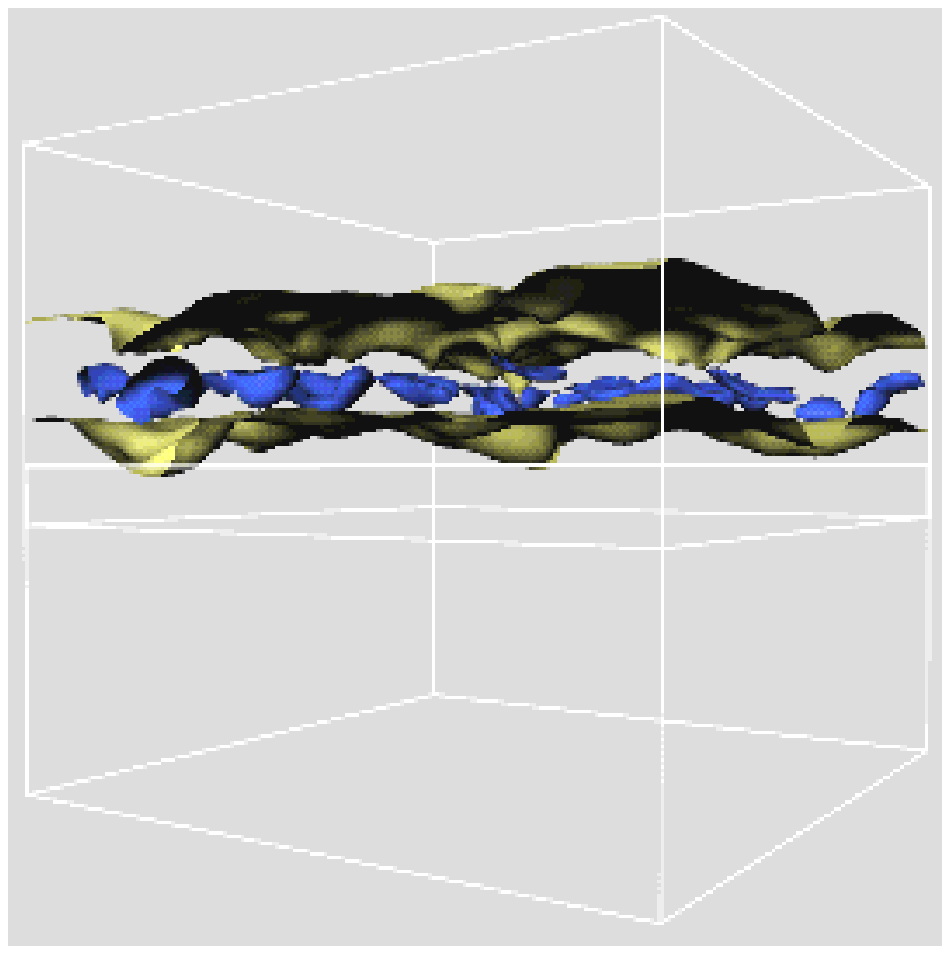}
\epsfxsize=7.0cm \epsfbox{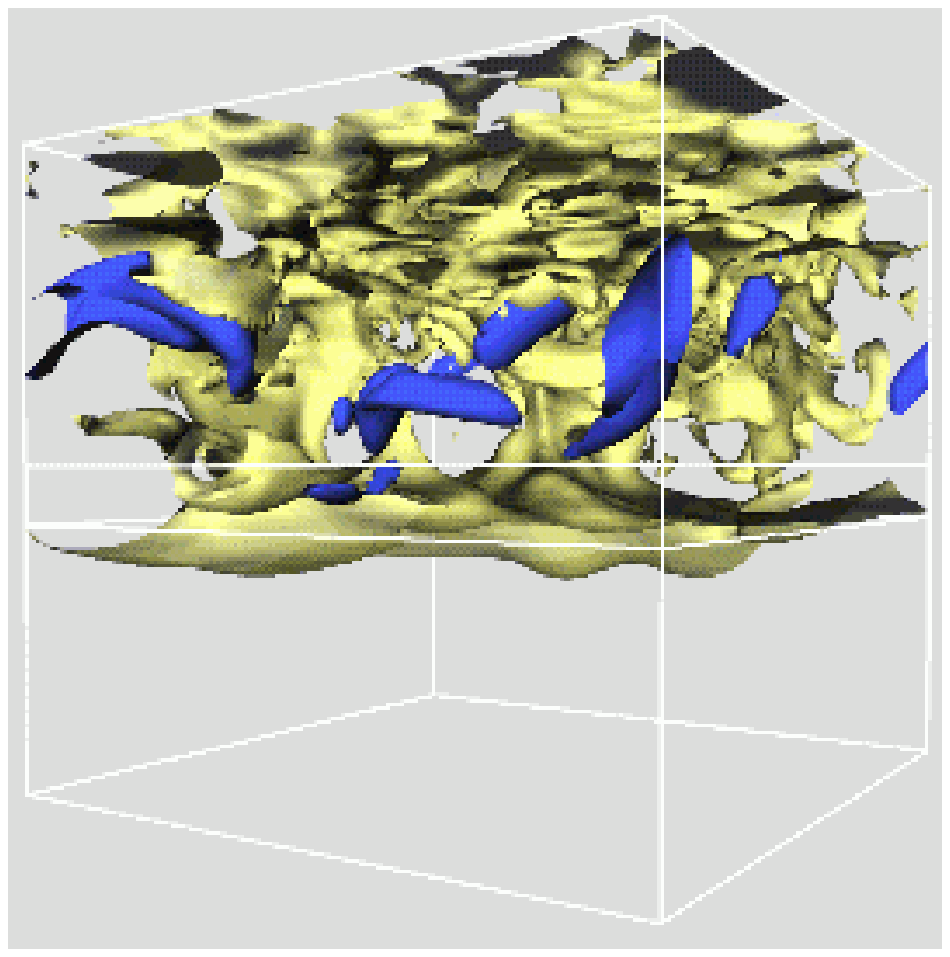}}
\vspace{0cm}}
\caption[]{Snapshot of magnetic field isosurfaces:
Light gray (yellow), low field strength --- dark grey (blue), 10 times
higher field strength (colors refer to color versions of the figure,
available at {\tt http://www.astro.ku.dk/$\sim$aake}).
Left: Early in the experiment. Right: At a late time. } \label{fig1}
\end{figure*}

\begin{figure}
\centering
\makebox[8.8cm]{
\vspace{0cm}
\hspace{0cm}\epsfxsize=8.8cm \epsfbox{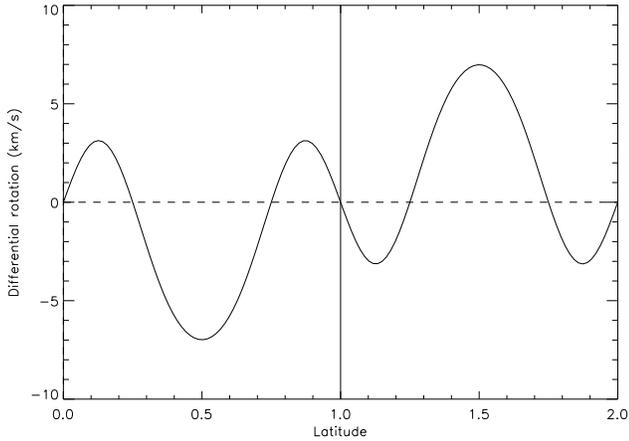}
\vspace{0cm} }
\caption[]{\small The background rotational velocity Eq.\ \ref{rotvel}
(in km/s) as a function of latitude (in units of $\pi$); the two
equators of the model are at the latitudes 0.5 and 1.5
(corresponding to $\theta = \pi/2$ and $\theta=3\pi/2$, respectively),
while the poles are at 0, 2 (north pole) and
1.5 (south pole). } \label{fig2}
\end{figure}

\subsection{Initial conditions}

The hydrodynamic part of the initial condition
is a snapshot from a well developed stage of a numerical model of the solar
convection zone.
The initial condition for the magnetic field is given by a unidirectional
(``poloidal'') sheet that is placed in the middle of the convection
zone.  The sheet is initially
in a state of isentropic pressure equilibrium with the surroundings.

The model has a high degree of stratification with
a density contrast of roughly $5\, 10^3$
in the convection zone alone.
The thickness of the undershoot layer in the model is approximately
equal to 0.8 ${\rm H}_{\rm P}$, where ${\rm H}_{\rm P}$ is the pressure
scale height at the bottom of the convection zone. This
is much larger than the helioseismological upper limits of 0.1 -- 0.2
${\rm H}_{\rm P}$ (see e.g.\ Christensen-Dalsgaard \cite{JCD+ea95}).
Scaling the extension of the undershoot
layer naively in proportion to the velocity (Hooke's law --- the
retarding force increases approximately linearly with distance),
the resulting undershoot thickness, of
the order of 0.01 ${\rm H}_{\rm P}$, falls well within
the upper limits from observations.

\subsection{Latitudinal shear}
\label{lat.shear}

Both experiments with and without differential rotation
in the convection zone were performed, with
the purpose of
illustrating the effects of radial and latitudinal shear and of the
Coriolis force on the magnetic field.
In the rotational cases background differential rotation and the Coriolis force
are included by using the force ${\bf F}_{\rm ext}$ in the equation
of motion (Eq.\ \ref{EoM})
\begin{equation}
{\bf F}_{\rm ext} = {\bf F}_{\rm rot} + {\bf F}_{\rm Coriolis}
\end{equation}
where ${\bf F}_{\rm Coriolis} = - 2\rho {\bf \Omega} \times {\bf u}$
is the Coriolis force and ${\bf F}_{\rm rot}$ is a force designed to
induce background rotation (see below).

The differential rotation that is implemented in the model may not be
identified as rotation around e.g.\ one of the horizontal axes.
Rather the mapping of the rotation is such that the rotation axis is
given by a vector in the ``meridional'' $(x,y)$ plane pointing towards the
north pole, i.e.\ a rotation vector
${\bf \Omega} = (\Omega_x,\Omega_y,0)$  given by
\begin{eqnarray}
\Omega_x & = & \Omega(x,\theta) \cos \theta,\\
\Omega_y & = & \Omega(x,\theta) \sin \theta,
\end{eqnarray}
where $\Omega=\Omega(x,\theta)$ is a fit to
the observed solar angular frequency
(from Dziembowski et al. \cite{Dziembowski+ea89})
and $\theta \in [0:2\pi],~ \theta(y)
= 2\pi y/L_y$ where $L_y$ is half the size of the box in the latitudinal
direction. This choice means that ${\bf \Omega}$ is
parallel to the x-axis (the radial direction) at the poles
and parallel to the y-axis (the latitudinal
-- or poloidal direction) at the equator.

An azimuthal background
velocity corresponding to the observed rotation is implemented by adding
a Newtonian term in the equation of motion:
\begin{equation}
 {\bf F}_{\rm rot} = - \frac{\rho}{\tau_R} ( u_z - u_z^0 ) \hat{{\bf z}},
\label{rotation}
\end{equation}
where $\tau_{\rm R}$ is the time scale of forcing
of the background rotation (one turn-over time has proved to be a good choice),
$\hat{{\bf z}}$ is the the
unit vector in the azimuthal direction, and $u_z^0=u_z^0(x,y)$ is the
background rotation speed given by
\begin{equation}
        u_z^0 = - (\Omega(x,\theta) - \Omega_0) R \sin \theta. \label{rotvel}
\end{equation}
where
$R$ is a characteristic radius in the convection zone and $\Omega_0$
is the angular rotation frequency evaluated at $\theta = \frac{\pi}{4}$.

The differential rotation is then measured relative to the latitude
$\pm 45^{\rm o}$. In the model that latitude corresponds to the position
at fractions of 0.25, 0.75, 1.25 and 1.75 of the latitudinal extension
of the
computational box counting the total length as $2 L_y$.
(see Fig.\ \ref{fig2}). 
At that latitude there is no radial shear since we set the internal solid
body rotation equal to the in situ azimuthal speed, so that
\[  \Omega = \left\{ \begin{array}{ll}
                \Omega(x, \theta ), & \mbox{$x \geq R$} \\
                \Omega_0,           & \mbox{$x < R$}
                \end{array}
        \right. \] 
At other latitudes the peak
of the radial shear is located in the undershoot layer.

The strength of the background rotation may also be changed by
varying the radius $R$ in Eq.\ \ref{rotvel}.
Values from 500 to 1400 Mm were used, the higher values in
combination with larger values of $\tau_R$, in order to still produce
the desired amplitude of the differential rotation.

\section{Results}

Several experiments with varying initial magnetic field
strengths, numerical resolutions, and upper boundary conditions were
performed.
First, results from simulations without rotational effects are
discussed (see Table \ref{pumpexperiments}), and then results
from a simulation including latitudinal shear are reported.
\begin{table*}[!htb]
\begin{center}
\begin{tabular}{|c|c|c|c|c|} \hline \hline
Experiment & Initial field & Plasma $\beta$ & Grid points & Top boundary \\ \hline
  [i] & Sub-equipartition   & $2\times 10^3$ & $78\times64\times64$ & Closed \\ \hline
 [ii] & Kinematic           & $2\times 10^5$ & $78\times64\times64$ & Closed \\ \hline
[iii] & Sub-equipartition   & $2\times 10^3$ & $63\times64\times64$ & Open   \\ \hline
 [iv] & Super-equipartition & $200$          & $63\times64\times64$ & Open   \\ \hline
  [v] & Super-equipartition & $20$           & $63\times64\times64$ & Open   \\ \hline
 \hline
\end{tabular}
\end{center}
\caption[]{\small Summary of five experiments not including rotational
effects.}
\label{pumpexperiments}
\end{table*}
Although some of the experiments initially had a maximum field strength of
the order of the formal equipartition value, they have been denoted
`sub-equipartition' in Table \ref{pumpexperiments} because the
distribution of the kinetic energy density $e_K$ was much broader than the
distribution of the magnetic energy density $e_M$, i.e.\ the peak
magnetic energy density was smaller than the peak kinetic energy
density even though the most likely value of $e_M$ was
similar to that of $e_K$.

A moderate number
of grid points were used in these experiments (a few experiments were also
ran at higher resolutions, up to $145\times128^2$, to check for resolution
effects).
The advantage of performing relatively small numerical experiments, is
that it is possible to perform a larger number of experiments, with
different setups and with a variety of parameter values.
Since it is impossible, with the limits of currently available computer
power to accurately reproduce solar conditions, it is necessary to
experiment with trade-offs between various constraints.

\subsection{The rotation-less case}

Fig.\ \ref{fig1} shows two sets of magnetic field isosurfaces at two
different instants in time for an experiment with a sub-equipartition magnetic
field (experiment [i], in Table \ref{pumpexperiments}):
The poloidal sheet of magnetic field initially
placed in the middle of the convection zone quickly starts to interact with
the convection (Fig.\ \ref{fig1}, left panel).
At a subsequent
time (Fig.\ \ref{fig1}, right panel) the magnetic field
more or less fills the whole volume of the convection zone and
penetrates into the stable layer below.

\begin{figure}
\makebox[8.8cm]{
\vspace{0cm}
\hspace{0cm}\epsfxsize=8.8cm \epsfbox{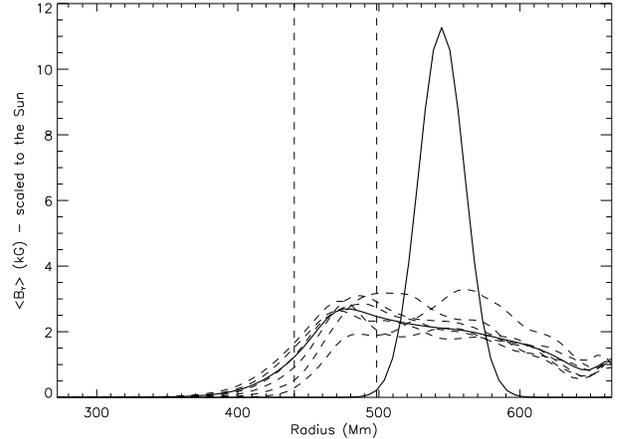}
\vspace{0cm} }
\caption{The horizontally averaged magnetic field $\langle
B_y \rangle$ as a function of radius for 7 different times
(experiment [i]):
$t=$ 0 (initial field --
solid curve), 3.3, 6.5, 9.8, 13.0, 16.3 (thin dashed curves),
and 19.5 (solid curve) turn-over times.
The approximate bottom and top of the undershoot layer are indicated by two
vertical dashed lines.}
\label{fig3}
\end{figure}

Fig.\ \ref{fig3} shows the horizontally averaged field at 7 equidistant
instants of time for experiment [i]:
The poloidal sheet is spread out, and the distribution of poloidal magnetic
flux settles to a characteristic distribution.  The highest (horizontally
averaged) poloidal flux density occurs in the overshoot layer, and in these
particular experiments a significant fraction of the total flux also resides
there.  In the real Sun, this fraction may be expected to be much smaller,
because the undershoot layer is much thinner there.  Given the shape of the
distribution, with no particular enhancement in the undershoot layer, it is
likely the Sun has a correspondingly smooth distribution,
with the majority of the poloidal flux residing inside the convection zone.

The `pumping effect' described in the above
takes place because of the topology of the over-turning
stratified convection: Of all the fluid parcels threaded by
magnetic field lines
in the initial state about half are initially ascending.  However, because
of the stratification, most of the ascending fluid parcels have to over-turn
and descend and most of these keep descending down to the bottom of the
convection zone (Stein \& Nordlund \cite{Stein+Nordlund89b}).
The fluid parcels drag the
threading field lines along and hence
an appreciable fraction of the field is transported downwards.
Fragments of the field that are caught in ascending flows are
advected upwards.  In the experiments with an open upper boundary some
of these fragments escape through the top of the computational domain
and flux is systematically lost
(see the discussion below).

Even in the cases of super-equipartition fields (e.g.\ experiments [iv]
and [v])
the magnetic field is pushed downwards in the initial phases
until it reaches the undershoot layer
where the kinetic energy flux decreases.
Because the overall magnetic flux decreases through flux loss at the surface
the magnetic field eventually enters a
state in which it is below equipartition.

\begin{figure*}
\centering
\makebox[14cm]{
\vspace{0cm}
\hbox{\hspace{0cm}\epsfxsize=7.0cm \epsfbox{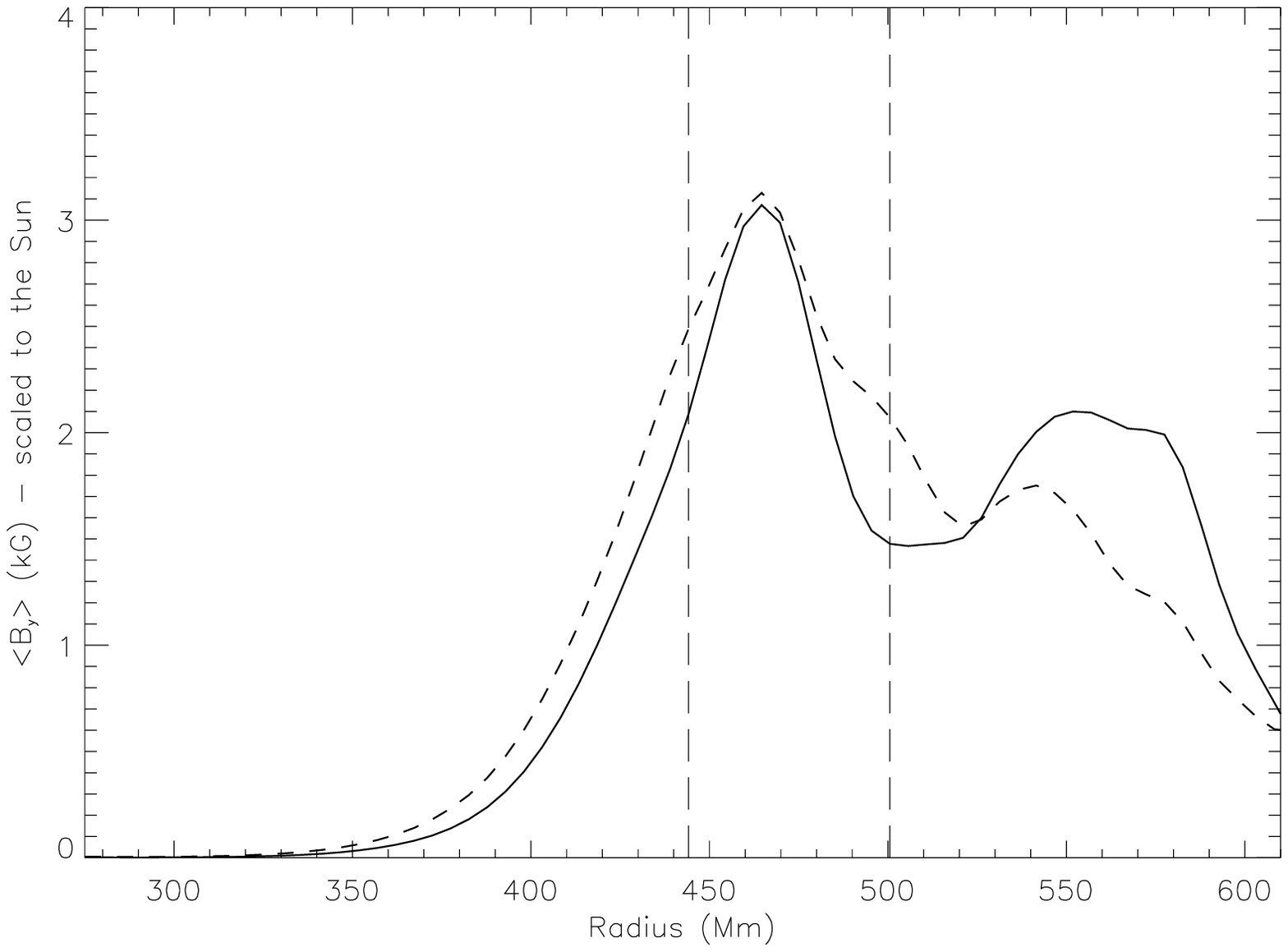}
\epsfxsize=7.0cm \epsfbox{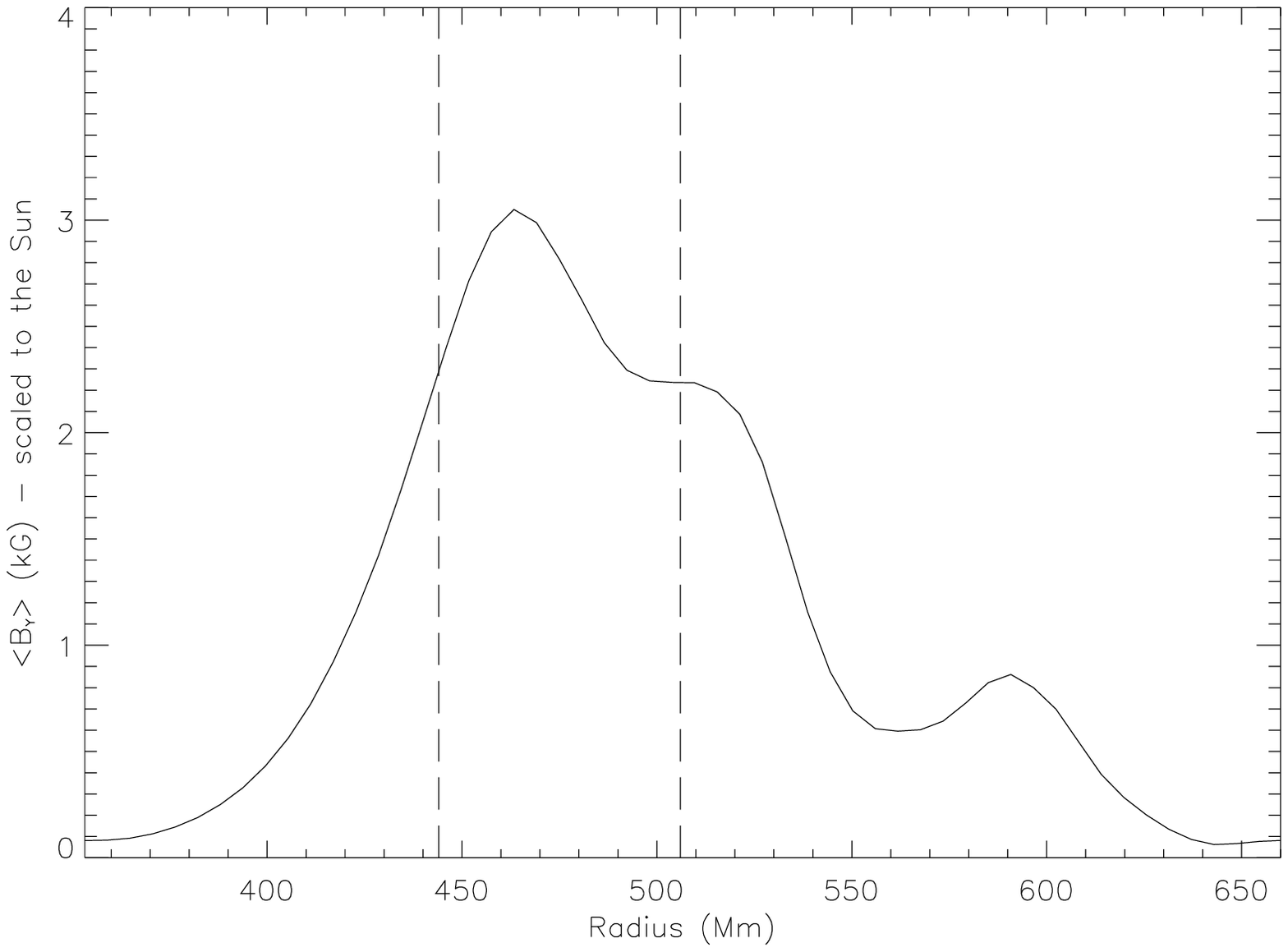}}
\vspace{0cm} }
\caption{Horizontal average of the poloidal $B_y$ field for two experiments
(left panel: solid line, experiment [i] and dashed line, experiment [ii])
that had
a closed upper boundary (after 25 and 30 turn-over times respectively),
and for experiment [iii] (right panel) that had an open upper
boundary (after 18 turn-over times).
The amplitudes of the curves have been scaled
to have the same maximum as experiment [i]
to compare to experiment [i].}
\label{fig4}
\end{figure*}
In the case of an initial field below equipartition
the  pumping effect is --- not surprisingly --- even more pronounced than for
the super-equipartition fields: Fig.\
\ref{fig4} shows the results for
three cases where the magnetic field was below equipartition
(experiments [i], [ii] and [iii])
and the dynamics thus were dominated by the convective
motions. The average magnetic field is distributed over the entire convection
zone, with maximum (horizontally averaged) flux density in the
undershoot layer.

The radial distribution of the magnetic flux present
in the convection zone
is not very different for the cases
with a closed upper boundary (left panel in Fig.\ 
\ref{fig4}) and the cases with an open upper boundary
(right panel in Fig.\ \ref{fig4}), but the total amount of magnetic flux
is rapidly reduced during the
first few turn-over times, when there is a significant loss of
poloidal flux through the upper boundary.
This does not influence the distribution of flux over depth much, however,
it only influences the amount of flux that is available for distribution.

The flux loss in the models with an open upper boundary is
strongly exaggerated in comparison to the Sun.
The real flux loss may be expected to be significantly smaller
than in the models with open upper boundaries discussed here: Much of
the weak ascending flux must over-turn rather than reach the solar
surface since it is embedded in a fluid of which only a tiny fraction
($\sim$ 0.1\% estimated from mass flux amplitudes)
reaches the solar surface.  Note that the magnetic field considered
is weak and incoherent and does not have sufficient buoyancy to
overcome the drag of the fluid motions.

A supplementary numerical experiment confirms the exaggeration
of the flux loss: Moving the open magnetic boundary four mesh
point upwards, corresponding to about 0.6 density scale heights,
reduces the flux loss by about 20\%.  The new boundary (at
$\rho = 2.5\times10^{-2}$ gcm$^{-3}$) is still over ten density
scale heights (sic) away from the real solar boundary (at $\rho \sim 3\times
10^{-7}$ gcm$^{-3}$).  Almost all of the flux that is still lost in the
modified experiment may thus in the case of the real Sun be expected
to turn over before reaching the solar surface.

\begin{figure*}[!htb]
\centering
\makebox[14cm]{
\epsfxsize=14.0cm
\epsfysize=11.0cm
\epsfbox{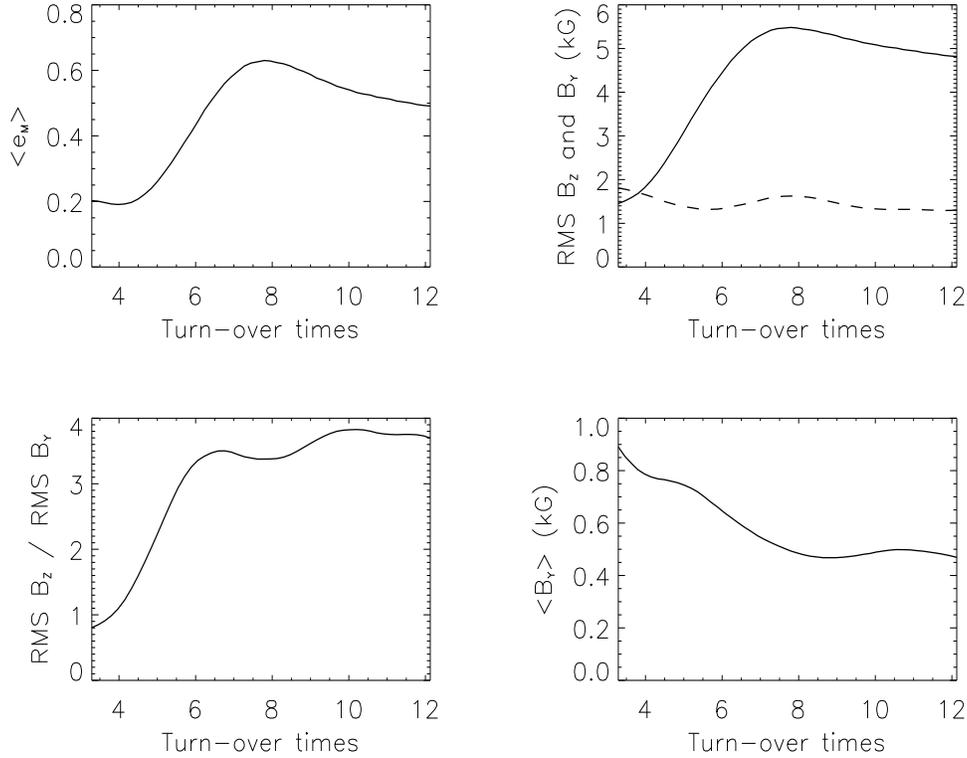} }
\caption[]{\small The case with latitudinal shear:
Four panels showing the average magnetic energy $\langle
e_M \rangle$, the RMS field strengths $B_z$ (toroidal field --- solid line) 
and $B_y$ (poloidal field --- dashed line), 
the ratio of toroidal to poloidal field, and the average poloidal
field $\langle B_y \rangle$ as functions of turn-over times.} \label{fig5}
\end{figure*}

\begin{figure}
\makebox[8.8cm]{
\vspace{0cm}
\hspace{0cm}\epsfxsize=8.8cm \epsfbox{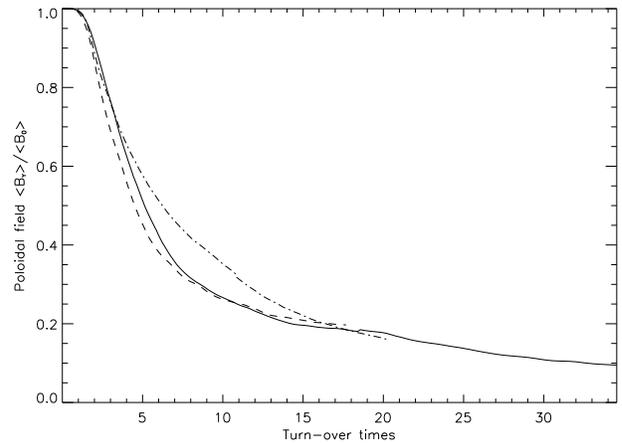}
\vspace{0cm} }
\caption{The normalized average poloidal field strength as function
of time in turn-over times for the experiments with an open upper
boundary condition:
(experiment [iii] dashed curve,
experiment [iv] solid curve and experiment [v] dashed-dotted curve).}
\label{fig6}
\end{figure}

Fig.\ \ref{fig6} shows the average poloidal field strength
for the experiments with an open upper boundary
(experiments [iii], [iv] and [v]). The relative reduction in the poloidal
flux by the escape of magnetic structures through the upper boundary
does not depend much on the field strength, for strengths up to
several times formal equipartition values.

The rate of flux loss does depend on the field strength but in
a rather counterintuitive way: The stronger
the initial field the smaller the initial reduction rate.
Naively one might expect that if the field strength is large
the higher buoyancy would make the field escape faster.
The explanation is that the over-turning of the fluid
is slowed down in the cases with a stronger initial poloidal flux sheet.
The fragmentation of the sheet is also slower in the case
with a stronger field.

The magnetic field in the convection zone
rapidly becomes very fragmented:
While
the field in the beginning is uniformly distributed in the sheet,
quickly a picture develops
where the field in the convection zone becomes
very intermittent, while the field that is
pumped into the undershoot layer is much more uniform.

The reason for this difference in topology between the magnetic field
in the convection zone and in the undershoot layer is that
once the magnetic field has been pumped down into
the undershoot layer it is less susceptible to fragmentation, since the
motions in the stable layer have a much smaller amplitude and are not as
systematic as the motions in the convection zone.

The degree of intermittency and fragmentation of the field in the convection
zone depends on the strength of the initial poloidal magnetic sheet.
The poloidal magnetic sheet is more stable towards
the initial fragmentation if its field is stronger since in that case
a larger force is needed to over-power the tension in the poloidal field lines.

\subsection{A case with latitudinal shear}
\label{rotation.ch}

In what follows, results are presented
for a particular simulation that includes shear and the Coriolis force
(through the force ${\bf F}_{\rm ext}$
in the equation of motion Eq.\ \ref{EoM}).
The physical size of this experiment was $381\times800\times800$ Mm
and the numerical resolution $69\times 135^2$ grid points.

The initial condition of a uniform poloidal sheet placed at a certain
depth may be said to be rather arbitrary. However, an arbitrary state
is both appropriate and useful in this case, since the system rather quickly enters
a state that is independent of the initial condition i.e.\
there is no internal long-time ``memory'' in the system.

\begin{figure}
\centering
\makebox[8.8cm]{
\vspace{0cm}
\hspace{0cm}\epsfxsize=8.8cm \epsfbox{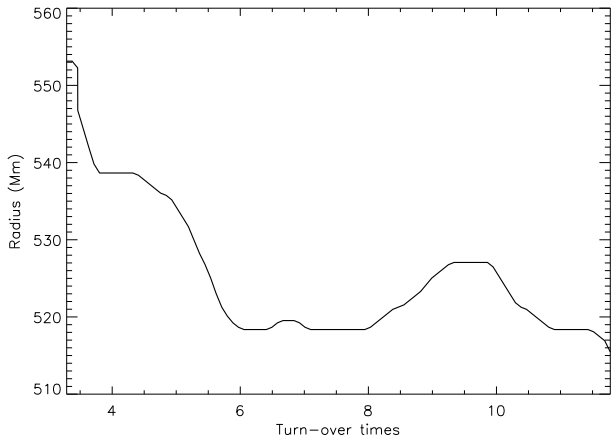}
\vspace{0cm} }
\caption[]{\small The depth of the location of the maximum magnetic
energy $\langle e_M \rangle$ as a function of turn-over times.}
\label{fig7}
\end{figure}

Initially the pumping effect ensures that an on the average poloidal field
is injected into the bottom undershoot layer, and after about
four turn-over times the poloidal flux density
has a maximum below the convection zone of the model.

After an initial transient process where the differential rotation
establishes itself and the over-turning
convection distorts the poloidal sheet, the system enters
a well-developed state, where the magnetic field displays the
structure of the differential rotation. At this point, the magnetic energy
and the toroidal field strength have increased to a significantly higher level
than their initial values (see the two top panels in Fig.\ \ref{fig5}),
and the rms toroidal field is larger than the rms poloidal field.  The average
poloidal flux has dropped to about half the initial value due to flux loss
at the surface (see the bottom two panels in Fig.\ \ref{fig5})
but this loss of magnetic flux through the upper boundary is
somewhat halted after the formation of a large-scale toroidal field.

At first the growth of the toroidal field is fast while the
background rotation increases toward the profile determined by Eq.\
\ref{rotvel} on the time scale $\tau_R$ (cf.\ Eq.\ \ref{rotation}).
When the
background rotation profile is fully attained by the fluid, the
toroidal magnetic field begins to increase linearly, with the rate of
increase given by the latitudinal shear and the original poloidal
field strength.
The structure of the magnetic field directly
reflects the latitudinal dependence of
the background azimuthal velocity field
as a result of the latitudinal shear.

\begin{figure}
\centering
\makebox[8.8cm]{
\epsfxsize=8.8cm
\epsfysize=8.8cm
\epsfbox{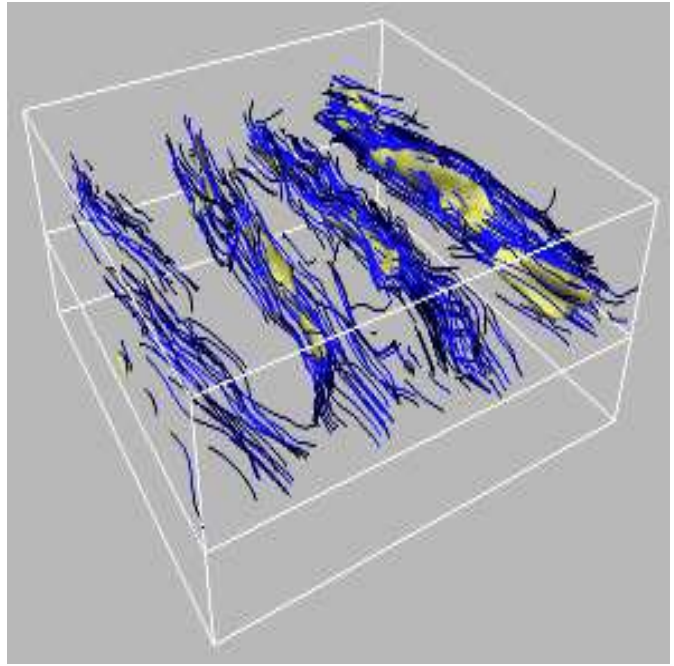} }
\caption[]{\small A snapshot of strong magnetic field lines (blue) and
magnetic field isosurfaces (yellow).} \label{fig8}
\end{figure}

In the well-developed state after the main flux loss has taken place,
the dominant magnetic field
is a strong toroidal flux system located near the bottom of
the convection zone, i.e.\ it is not pumped into the undershoot layer
but it does not escape from the convection zone either: The
``center of gravity'' of the magnetic field is
above the bottom of the convection zone (see Fig.\ \ref{fig7}).

\begin{figure*}
\centering
\makebox[14cm]{
\epsfxsize=7.0cm
\epsfysize=10.0cm 
\epsfbox{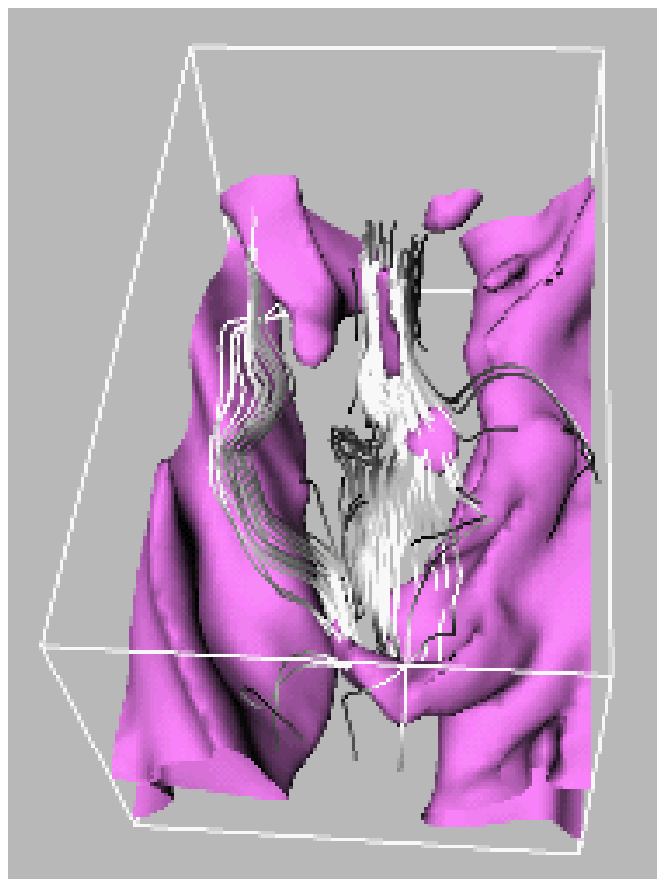} 
\epsfxsize=7.0cm
\epsfysize=10.0cm
\epsfbox{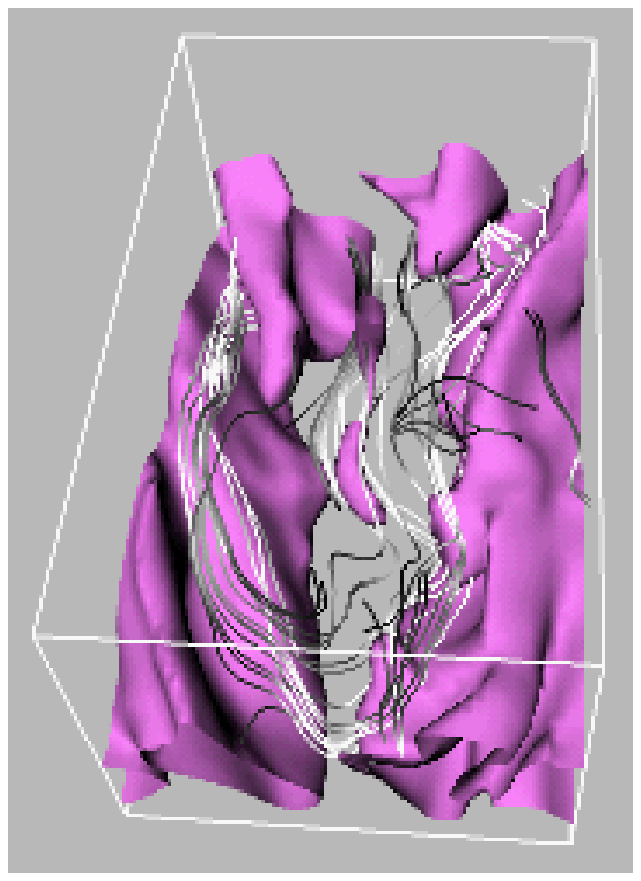} }
\caption[]{\small A view of one quarter of the box, centered at one of the
equators, where a magnetic structure is emerging through the surface and
is moving towards the equator. Field
lines in white and grey (purple) isosurfaces of magnetic field strength.
The snapshot to the right is taken $\sim$ half a turnover time later
than the one on the left. } \label{fig9}
\end{figure*}

Fig.\ \ref{fig8} shows a view of the toroidal flux system:
Four toroidal flux streets
are formed on each side of the equators, in the regions where there
is a maximum shear.

The magnetic field just below the convection zone rotates more slowly
than the field just above. This is a mechanism that may twist the
magnetic field lines that connect across the equator regions.

As the initial poloidal sheet fragments while being wound up, toroidal
flux structures leave the convection zone through the open upper boundary.
In a rotating frame of reference vertical motions lead to horizontal
motions through the action of the Coriolis force. One special case is the
rise of flux loops that are rotated so that they emerge tilted with respect
to the latitudinal circle, and another is the excitation of meridional
circulation, as a result of the transport of angular momentum:
In the simulation both of these effects are found.

The surface of the Sun is an extremely important place: Not only may flux be
lost there, but the surface constitutes a
``reconnection region'' where vertical field lines are effectively
``cut over'', and where the remaining ``stubs'' may be advected passively
by the horizontal surface motions.  Field lines of course actually continue
into the corona and either connect back into the photosphere or connect
out into interplanetary space.  However, reconnection is observed to proceed
so rapidly in the corona and the mass density is so low there, that from
the point of view of the sub-surface dynamics the connections above the
surface are of little importance.

Fig.\ \ref{fig9} shows two panels with a small section of the
box, with  snapshots of the magnetic field, looking down along the two
toroidal flux streets.  A magnetic flux structure is sticking out through
the surface, and is moving towards the equator between the two
toroidal flux streets, but at the same time
it is connected through a subsurface structure to one of them.
As a result of the drift of the surface-structure, the magnetic field
lines that make up the subsurface structure have a tendency to become
inclined with respect to the toroidal street to which it connect, and hence
a poloidal field component is formed, with opposite polarity relative
to the original one.

\section{Discussion and conclusions}

The main conclusions that may be drawn from the study of the interaction
of a magnetic sheet with stratified over-turning convection in the absence
of rotation are the following:
\begin{itemize}
\item
Stratified convection induces a strong tendency to transport magnetic flux
downwards.
\item
The distribution of horizontally averaged magnetic flux peaks in the
undershoot layer, but the bulk of the flux is in the convection
zone.
\item
The magnetic field that resides in the undershoot layer is considerably
less fragmented than the magnetic field in the bulk of the convection
zone.
\item
Unless an open upper boundary is placed sufficiently close to the
actual solar surface, i.e.\ at sufficiently low density, there may
be a substantial (and exaggerated) loss of flux through the open
boundary as flux is carried around by the over-turning convection.
\item
The transport properties (both the downwards transport and the
surface flux loss) are quite robust and field strengths well in excess
of formal equipartition are needed to change the distribution and
rates significantly.
\end{itemize}
Additionally, the results of the simulations including latitudinal shear
offer the following conclusions:
\begin{itemize}
  \item The system rapidly forgets the initial condition and distributes
        the flux in a generic vertical distribution.
  \item The latitudinal shear of the differential rotation shapes the
        magnetic field and creates strong toroidal flux streets
        located at mid latitudes.
  \item While magnetic flux indeed is pumped into the undershoot layer,
        the center of gravity of the magnetic field is above the bottom
        of the convection zone.
  \item Magnetic field lines and flux structures penetrating the upper
        boundary move passively according to the surface motion.
\end{itemize}

Where do these results leave the ``storage problem'', i.e.\ the problem
of explaining how the magnetic field can remain stored while
being amplified by differential rotation?

The toroidal magnetic field in these simulations reach peak field strengths
of a several tens of kG (scaled to the Sun).  The peak field strengths
occur near the bottom, but still inside, the convection zone. It is
conceivable that emerging flux regions form when buoyancy finally becomes
dominant, and that this occurs at field strengths of the order of 100 kG,
as has been deduced from emergence and tilt patterns by several investigations.
Because of numerical limitations we were not study that process with the
current series of experiments---future experiments with higher resolution
and even larger density contrasts are needed here.

Besides offering a clue to the operation of the solar dynamo the results
presented here may also contribute to the understanding of magnetic field
generation in other late type stars.  For example; the fact that no
undershoot storage is available in the case of
magnetically active fully convective M dwarf stars
(e.g.\ Chabrier \& Baraffe \cite{Chabrier+Baraffe97})
is generally considered to be a problem (see e.g.\
Allard et al. \cite{Allard+ea97} and
K\"{u}ker \& R\"{u}diger \cite{Kueker+Ruediger99}) --- this ceases
to be a problem in the scenario presented here.

We note that, even though the loss of magnetic flux at the upper boundary is
exaggerated in the models presented here, such a loss is certainly a real
effect that is important to include, since the Sun is known to loose a
considerable amount of toroidal magnetic flux during an activity cycle
(e.g.\ van Ballegooijen \cite{Ballegooijen1998}).

Lastly we find it important to emphasize that the visualization
of ``emerging'' flux structures and field lines (Fig.\ \ref{fig9})
illustrates a mechanism first pointed out
by van Ballegooijen (\cite{vanBalleprivate95})
that may be crucial for the reversal of the poloidal
field: Flux structures that rise and penetrate the surface effectively
results in the field lines being ``cut'' at the surface, with the
leading polarity tending to drift towards the equator, and the following
polarity tending to drift towards the pole.  The result is that subsurface
connections between the following polarity of one emerging structure and
the leading polarity of another trailing structure may become tilted in
the sense opposite to the tilt associated with the normal
winding of the field.  When such reversed tilts are caught by the
differential rotation, they will effectively lead to ``unwinding''
and reversal of the poloidal field component.  This again is an important
topic for future studies.

\begin{acknowledgements}
This work was supported in part by the Danish Research Foundation,
through its establishment of the Theoretical Astrophysics Center.
Computing time at the UNI$\bullet$C computing center
was provided by the  Danish Natural Science Research Council.
SBFD acknowledges support through an EC-TMR grant to the European
Solar Magnetometry Network.
\end{acknowledgements}


\begin{thebibliography}{}

\bibitem[1997]{Allard+ea97}
Allard, F., Hauschildt, P.H., Alexander, D.R., Starrfield, S.,
1997, Ann.\ Rev.\ A\&A 35, 137

\bibitem[1983]{Arter1983}
Arter, W., 1983, J.Fluid Mech. 132, 25

\bibitem[1982]{Arter+ea82}
Arter, W., Proctor, M. R.E., Galloway, D., 1982, MNRAS 201, 57P

\bibitem[1996]{Brandenburg+96jfm}
Brandenburg, A., Jennings, R.L., Nordlund, A., Rieutord, M., Stein, R.F.,
  Tuominen, I. 1996, JFM 306, 325

\bibitem[1990]{Brandenburg+ea90}
Brandenburg, A., Nordlund, {\AA}., Pulkkinen, P., Stein, R.F., Tuominin,
  I., 1990, A\&A 232, 277

\bibitem[1999]{Chabrier+Baraffe97}
Chabrier, G., Baraffe, I., 1997, A\&A 327, 1039

\bibitem[1995]{Choudhuri+ea95}
Choudhuri, A.R., M., S., Dikpati, M., 1995, A\&A 303, L29

\bibitem[1995]{JCD+ea95}
Christensen-Dalsgaard, J., Monteiro, M.J.P.F.G, Thompson, M.J., 1995,
MNRAS 276, 283

\bibitem[1980]{Drobyshevski+ea80}
Drobyshevski, E., Lokesnikova, E.N., Yuferev, V., 1980,
J.Fluid Mech. 101, 65

\bibitem[1974]{Drobyshevski+ea74}
Drobyshevski, E., Yuferev, V., 1974, J.Fluid Mech. 65, 38

\bibitem[1989]{Dziembowski+ea89}
Dziembowski, W.A., Goode, P.R., Libbrecht, K.G., 1989, ApJ 337, L53

\bibitem[1983]{Galloway+ea83}
Galloway, D., Proctor, M. R.E., 1983, Geophy. \& Astroph. Fluid Dyn. 24, 109

\bibitem[1997]{Galsgaard+ea97}
Galsgaard, K., Nordlund, {\AA}, 1997, Journ. Geoph. Res. 102, 219

\bibitem[1988]{Hurlburt+ea88}
Hurlburt, N.E., Toomre, J., 1988, ApJ 327, 920

\bibitem[1994]{Hurlburt+ea94}
Hurlburt, N.E., Toomre, J., Massaguer, J., Zahn, J.-P., 1994,
ApJ 421, 245

\bibitem[1979]{Hyman1979}
Hyman, J.M. 1979,
Adv. in Comp. Meth. for PDE's---III,
R. Vichnevetsky, R. S. Stepleman (eds.)
Publ. IMACS, p. 313

\bibitem[1992]{Jennings+ea92}
Jennings, R., Brandenburg, A., Nordlund, {\AA}., Stein, R.F., 1992,
MNRAS 259, 465

\bibitem[1996]{Johns-Krull+Valenti96}
Johns-Krull, C.M., Valenti, J.A., 1996, ApJ Letters 459, L95

\bibitem[1999]{Kueker+Ruediger99}
K\"{u}ker, M., R\"{u}diger, G., 1999, In: A.\ Ferriz-Mas, M.\ Nunez
(eds.), Stellar dynamos: Non-linearity and chaotic flows. ASP 178, p.87 

\bibitem[1998]{Mcleod98}
Mcleod, A.D., 1998, In S.M. Miyama et al.\ (ed.), Numerical Astrophysics, p.331

\bibitem[1992]{Nordlund+92}
Nordlund, A., Brandenburg, A., Jennings, R.L., Rieutord, M., Roukolainen, J.,
  Stein, R.F., Tuominen, I., 1992, ApJ 392, 647

\bibitem[1994]{Nordlund+94a}
Nordlund, A., Galsgaard, K., Stein, R.F., 1994,
In R.J. Rutten, C.J. Schrijver (eds.), Solar Surface Magnetic Fields
NATO ASI Series 433

\bibitem[1975]{Parker1975a}
Parker, E.N., 1975, ApJ 202, 523

\bibitem[1993]{Parker1993}
Parker, E.N., 1993, ApJ 408, 707

\bibitem[1989]{Rutten+ea89}
Rutten, R.G.M., Schrijver, C.J., Zwaan, C., Duncan, D.K.,
Mewe, R., 1989, A\&A 219, 239

\bibitem[1987]{Schrijver+Rutten87}
Schrijver, C.J., Rutten, R.G.M., 1987, A\&A 117, 143

\bibitem[1989]{Stein+Nordlund89b}
Stein, R.F., Nordlund, A., 1989, ApJ 342, L95

\bibitem[1998]{Tobias+ea98}
Tobias, S., Brummel, N., Clune, T., Toomre, J., 1998,
ApJ 502, L177

\bibitem[1995]{vanBalleprivate95}
van Ballegooijen, A.A., 1995, Private Communication

\bibitem[1998]{Ballegooijen1998}
van Ballegooijen, A.A., 1998, in Synoptic Solar Physics 140:
Kluwer Academic Publishers, p.17

\end{thebibliography}
\end{document}